\documentclass[11pt]{article}

\usepackage{graphicx}
\usepackage{amssymb}
\usepackage{amsthm}
\usepackage{amsmath}
\usepackage{tikz}
\usepackage{color}

\usepackage{tikz}
\usetikzlibrary{snakes,arrows,shapes}

\newtheorem{thm}{Theorem}

\newtheorem{lem}[thm]{Lemma}

\newtheorem{obs}[thm]{Observation}
\theoremstyle{definition}

\newcommand{\pig}{\mathrm{pig}}

\usepackage{subfig}
\usepackage{geometry}
\usepackage[colorlinks=true,urlcolor=black,linkcolor=black,citecolor=black]{hyperref}
\usepackage{algorithm}
\usepackage[noend]{algorithmic}
\algsetup{linenosize=\footnotesize,linenodelimiter=\empty}

\title{A Simple Characterization of the Minimal Obstruction Sets for Three-State Perfect Phylogenies}
\author{
	\normalsize Brad Shutters, David Fern\'{a}ndez-Baca\\
	\small Department of Computer Science, Iowa State University\\
	\small \href{mailto:shutters@iastate.edu,fernande@cs.iasate.edu}{\{shutters,fernande\}@iastate.edu}
	}
\date{}
\begin{document}
\maketitle
\thispagestyle{empty}
\begin{abstract}
Lam, Gusfield, and Sridhar (2009) showed that a set of three-state characters has a perfect phylogeny if and only if every subset of three characters has a perfect phylogeny. 
They also gave a complete characterization of the sets of three three-state characters that do not have a perfect phylogeny.
However, it is not clear from their characterization how to find a subset of three characters that does not have a perfect phylogeny without testing all triples of characters.
In this note, we build upon their result by giving a simple characterization of when a set of three-state characters does not have a perfect phylogeny that can be inferred from testing all pairs of characters.
\end{abstract}

\section{Introduction}

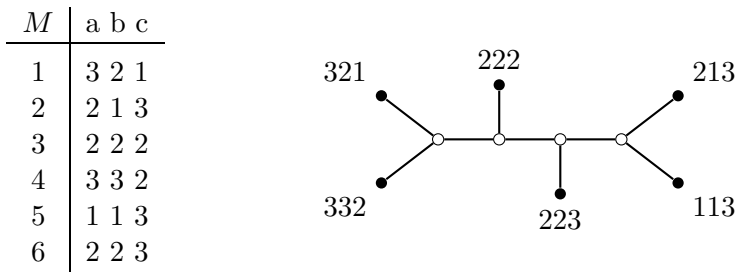
\begin{figure}[!h]
\centering
\begin{minipage}{1.55in}
	\centering
	\begin{tabular}{c|c}
	$M$ & a b c \\
	\hline
	\vspace{-9pt}\\
	1 & 3 2 1\\
	2 & 2 1 3\\
	3 & 2 2 2\\
	4 & 3 3 2\\
	5 & 1 1 3\\
	6 & 2 2 3
	\end{tabular}
\end{minipage}
\begin{minipage}{3in}
	\centering
	\begin{tikzpicture}[scale=.58]
	\node (n1) at (-1.3,1) [circle,inner sep=1.5,fill=black,label=above left:321] {};
	\node (n2) at (-1.3,-1) [circle,inner sep=1.5,fill=black,label=below left:332] {};
	\node (n3) at (0,0) [circle,inner sep=1.5,draw=black] {};
	\node (n4) at (1.4,0) [circle,inner sep=1.5,draw=black] {};
	\node (n5) at (1.4,1.25) [circle,inner sep=1.5,fill=black,label=above:222] {};
	\node (n6) at (2.8,0) [circle,inner sep=1.5,draw=black] {};
	\node (n7) at (2.8,-1.25) [circle,inner sep=1.5,fill=black,label=below:223] {};
	\node (n8) at (4.2,0) [circle,inner sep=1.5,draw=black] {};
	\node (n9) at (5.5,1) [circle,inner sep=1.5,fill=black,label=above right:213] {};
	\node (n10) at (5.5,-1) [circle,inner sep=1.5,fill=black,label=below right:113] {};
	\path[-] (n1) edge[thick] (n3);
	\path[-] (n2) edge[thick] (n3);
	\path[-] (n3) edge[thick] (n4);
	\path[-] (n4) edge[thick] (n5);
	\path[-] (n4) edge[thick] (n6);
	\path[-] (n6) edge[thick] (n7);
	\path[-] (n6) edge[thick] (n8);
	\path[-] (n8) edge[thick] (n9);
	\path[-] (n8) edge[thick] (n10);
	\end{tikzpicture}
\end{minipage}
\caption{Example 3-state perfect phylogeny for input matrix $M$.}
\label{fig3PPex}
\end{figure}

The {\em $k$-state perfect phylogeny problem} is one of the classic decision problems in computational biology. The input is an $n$ by $m$ matrix $M$ of integers from the set $\{1, \ldots, k\}$. 
We call a row of $M$ a {\em taxon} (plural {\em taxa}), a column of $M$ a {\em character}, and a value in column $c$ of $M$ a {\em state} of character $c$. 
A {\em perfect phylogeny} for $M$ is an undirected tree $t$ with $n$ leaves each labeled by a distinct taxon of $M$ in such a way that, for each character $c$ and each pair $i,j$ of states of $c$, the minimal subtree of $t$ containing all the leaves labeled by a taxon with state $i$ for character $c$ is node-disjoint from the minimal subtree of $t$ containing all the leaves labeled by a taxon with state $j$ for character $c$. The $k$-state perfect phylogeny problem is to decide whether $M$ has a perfect phylogeny. If $M$ has a perfect phylogeny, we say that the characters in $M$ are {\em compatible}, otherwise they are {\em incompatible}.
See \cite{Baca2001a, Semple2003a} for more on the perfect phylogeny problem.
See Figure \ref{fig3PPex} for an example.

If the number of states of each character is unbounded (so $k$ can grow with $n$), then the perfect phylogeny problem is NP-complete \cite{Bodlaender1992a,Steel1992a}. However, if the number of states of each character is fixed, the perfect phylogeny problem is solvable in  $O(m^2n)$ (in fact, linear time for $k=2$) \cite{Gusfield1991a, Dress1992a, Kannan1994a, Agarwala1994a, Kannan1997a}. Each of these algorithms can also construct a perfect phylogeny for $M$ if one exists. However, since every subset of a compatible set of characters is itself compatible, if no perfect phylogeny exists for $M$, there must be some minimal subset of the characters of $M$ that does not have a perfect phylogeny. We call such a set a {\em minimal obstruction set} for $M$. None of the above mentioned algorithms output a minimal obstruction set when there is no perfect phylogeny for $M$.

If the characters in $M$ are two-state characters, then $M$ has a perfect phylogeny if and only if the characters in $M$ are pairwise compatible. Hence, a minimal obstruction set for $k=2$ is of cardinality two \cite{Buneman1974a,Meacham1983a,Steel1992a,Estabrook1977a}.
A recent breakthrough by Lam, Gusfield, and Sridhar \cite{Lam2009a} shows that if the characters in $M$ are three-state characters, then any minimal obstruction set for $M$ has cardinality at most three.
It is conjectured that given an input matrix $M$ of $k$-state characters, there exists a function $f(k)$ such that $M$ has a perfect phylogeny if and only if every subset of $f(k)$ characters of $M$ has a perfect phylogeny \cite{Fitch1975a,Johnson1976a,Fitch1977a,Meacham1983a,Gusfield1991a,Lam2009a,Habib2011a}.
From the discussion above, it follows that $f(2)=2$ and $f(3)=3$. Recent work of Habib and To \cite{Habib2011a} shows that $f(4) \ge 5$.

If the characters in $M$ are $k$-state characters and the cardinality of a minimal obstruction set for $M$ is bounded above by $f(k)$, then it is preferable to have a test for the existence of such an obstruction set that does not require testing all subsets of $f(k)$ characters in $M$, and ideally one that can be inferred from testing all  pairs of the characters in $M$. Since we can decide if $M$ has a perfect phylogeny in $O(m^2n)$ time, and construct a perfect phylogeny in such a case, we should hope to also output a minimal obstruction set in $O(m^2n)$ time when a perfect phylogeny for $M$ does not exist.
 
Here, we will focus on the three-state perfect phylogeny problem. Hence, we restrict $M$ to be an $n$ by $m$ matrix of integers from the set $\{1,2,3\}$. 
We build upon the work of Lam, Gusfield, and Sridhar \cite{Lam2009a} who showed that if $M$ does not have a perfect phylogeny, then $M$ has an obstruction set of cardinality at most three.
They also gave a complete characterization of the minimal obstruction sets of cardinality three. 
However, it is not clear from their characterization how to find such an obstruction set without independently testing all triples of characters in $M$, requiring $O(m^3n)$ time.
In this note, we remedy this situation by giving a simple characterization of when a set of three-state characters does not have a perfect phylogeny that can be inferred from testing all pairs of characters in $M$.
This leads to a $O(m^2n)$ time algorithm to find an obstruction set when $M$ does not have a perfect phylogeny. If $M$ does admit a perfect phylogeny, then any of the above mentioned algorithms can be used to construct a perfect phylogeny for $M$ in $O(m^2n)$ time.

\section{Preliminaries}

\subsection{Perfect Phylogenies and Partition Intersection Graphs}

\begin{figure}[!h]
	\centering
	\begin{tikzpicture}[scale=.58]
	\node (a1) at (0,0) [circle,inner sep=1.5,draw=black,label=left:$a_1$] {};
	\node (a2) at (0,-2) [circle,inner sep=1.5,draw=black,label=left:$a_2$] {};
	\node (a3) at (0,-4) [circle,inner sep=1.5,draw=black, label=left:$a_3$] {};
	\node (b1) at (2,2) [circle,inner sep=1.5,draw=black, label=above:$b_1$] {};
	\node (b2) at (4,2) [circle,inner sep=1.5,draw=black,label=above:$b_2$] {};
	\node (b3) at (6,2) [circle,inner sep=1.5,draw=black, label=above:$b_3$] {};
	\node (c1) at (8,-2) [circle,inner sep=1.5,draw=black,label=right:$c_1$] {};
	\node (c2) at (8,0) [circle,inner sep=1.5,draw=black, label=right:$c_2$] {};
	\node (c3) at (8,-4) [circle,inner sep=1.5,draw=black,label=right:$c_3$] {};
	\path[-] (a3) edge (b2);
	\path[-] (a3) edge (b3);
	\path[-] (a2) edge (b1);
	\path[-] (a2) edge (b2);
	\path[-] (a1) edge (b1);
	\path[-] (b1) edge (c3);
	\path[-] (b2) edge (c1);
	\path[-] (b2) edge (c2);
	\path[-] (b2) edge (c3);
	\path[-] (b3) edge (c2);
	\path[-] (a3) edge (c1);
	\path[-] (a3) edge (c2);
	\path[-] (a2) edge (c2);
	\path[-] (a2) edge (c3);
	\path[-] (a1) edge (c3);
	\end{tikzpicture}
	\caption{Partition intersection graph of the matrix $M$ from Figure \ref{fig3PPex}.}
	\label{fig3PPpig}
\end{figure}
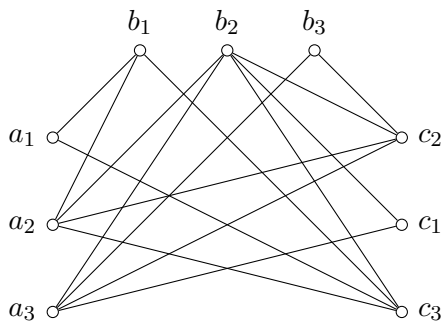

The {\em partition intersection graph} of $M$, denoted $\pig(M)$, is the graph that has a vertex ${c_i}$ for each character $c$ and each state $i$ of $c$, and an edge between two vertices $c_i$ and $d_j$ precisely if there is a taxon that has both state $i$ for character $c$ and state $j$ for character $d$. Note that there can be no edges between two vertices of the same character. See Figure \ref{fig3PPpig} for an example. In this section we give a brief overview of some known results relating three-state perfect phylogenies to partition intersection graphs.

A graph $G$ is  {\em triangulated} if and only if there are no induced chordless cycles of length four or greater. A {\em proper triangulation} of $\pig(M)$ is a triangulated supergraph of $\pig(M)$ such that each edge is between vertices of different characters.

\begin{thm}[Buneman \cite{Buneman1974a}, Meacham \cite{Meacham1983a}, Steel \cite{Steel1992a}]
There is a perfect phylogeny for $M$ if and only if $\pig(M)$ has a proper triangulation.
\end{thm}

For a subset $C=\{c_1,\ldots,c_j\}$ of the characters in $M$, we write $M[c_1,\ldots,c_j]$ to denote $M$ restricted to the columns in $C$. 
We say that $M$ is {\em pairwise compatible} if, for every pair $a,b$ of characters in $M$, there is a perfect phylogeny for $M[a,b]$.

\begin{thm}[Estabrook and McMorris \cite{Estabrook1977a}]\label{thmPairCompatible}
Let $a$ and $b$ be two characters of $M$. Then $M[a,b]$ has a perfect phylogeny if and only if $\pig(M[a,b])$ is acyclic.
\end{thm}

\begin{thm}[Lam, Gusfield, and Sridhar \cite{Lam2009a}]\label{thm3PP}
$M$ has a perfect phylogeny if and only if, for every three characters $a,b,c$ in $M$, $M[a,b,c]$ has a perfect phylogeny.
\end{thm}

If three of the characters are incompatible, then either they are not pairwise compatible, or, as the following theorem shows, the edges of their partition intersection graph is a superset (up to renaming of states) of one of a collection of ``forbidden'' edge sets.
%

\begin{thm}[Lam, Gusfield, and Sridhar \cite{Lam2009a}]\label{thm3StateGametes}
Let $M$ be pairwise compatible.
Then, a triple $\{a,b,c\}$ of characters in $M$ is a minimal obstruction set if and only if (under possibly renaming states) $\pig(M[a,b,c])$ contains all of the edges of one of graphs of Figure \ref{figObstructions}.
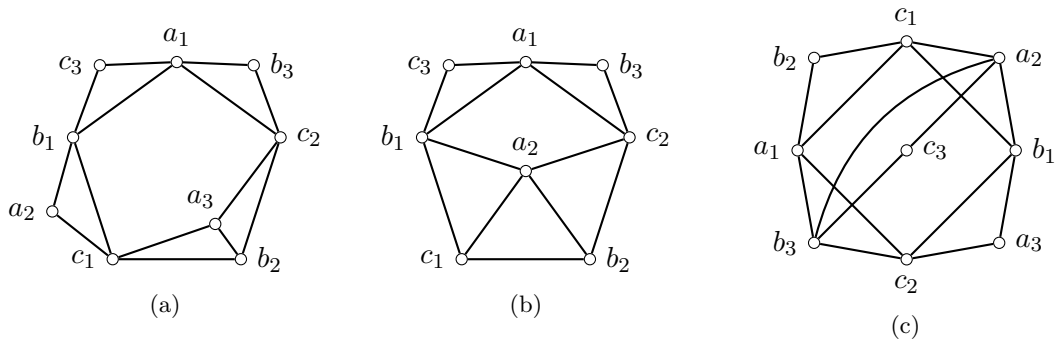
\begin{figure}[!h]
\centering
\begin{minipage}{1.9in}
	\subfloat[]{\label{figObstructions_1}
		\begin{tikzpicture}[scale=.58]
		\node (a1) at (90:2.5) [circle,inner sep=1.5,draw=black,label=above:$a_1$] {};
		\node (b1) at (162:2.5) [circle,inner sep=1.5,draw=black,label=left:$b_1$] {};
		\node (c1) at (234:2.5) [circle,inner sep=1.5,draw=black,label=left:$c_1$] {};
		\node (b2) at (306:2.5)  [circle,inner sep=1.5,draw=black,label=right:$b_2$] {};	
		\node (c2) at (18:2.5)  [circle,inner sep=1.5,draw=black,label=right:$c_2$] {};
		\node (b3) at (54:3) [circle,inner sep=1.5,draw=black,label=right:$b_3$] {};
		\node (c3) at (126:3) [circle,inner sep=1.5,draw=black,label=left:$c_3$] {};
		\node (a2) at (198:3) [circle,inner sep=1.5,draw=black,label=left:$a_2$] {};
		\node (a3) at (306:1.5) [circle,inner sep=1.5,draw=black,label=above:$a_3$\ \ \ \ ] {};
		\path[-] (a1) edge[thick] (b1);
		\path[-] (b1) edge[thick] (c1);
		\path[-] (c1) edge[thick] (b2);
		\path[-] (b2) edge[thick] (c2);
		\path[-] (c2) edge[thick] (a1);
		\path[-] (a1) edge[thick] (b3);
		\path[-] (c2) edge[thick] (b3);
		\path[-] (a1) edge[thick] (c3);
		\path[-] (b1) edge[thick] (c3);
		\path[-] (b1) edge[thick] (a2);
		\path[-] (c1) edge[thick] (a2);
		\path[-] (c1) edge[thick] (a3);
		\path[-] (b2) edge[thick] (a3);
		\path[-] (c2) edge[thick] (a3);
		\end{tikzpicture}
	}
\end{minipage}
\begin{minipage}{1.9in}
	\subfloat[]{\label{figObstructions_2}
		\begin{tikzpicture}[scale=.58]
		\node (a1) at (90:2.5) [circle,inner sep=1.5,draw=black,label=above:$a_1$] {};
		\node (b1) at (162:2.5) [circle,inner sep=1.5,draw=black,label=left:$b_1$] {};
		\node (ca) at (234:2.5) [circle,inner sep=1.5,draw=black,label=left:$c_1$] {};
		\node (b2) at (306:2.5)  [circle,inner sep=1.5,draw=black,label=right:$b_2$] {};	
		\node (c2) at (18:2.5)  [circle,inner sep=1.5,draw=black,label=right:$c_2$] {};
		\node (b3) at (54:3) [circle,inner sep=1.5,draw=black,label=right:$b_3$] {};
		\node (c3) at (126:3) [circle,inner sep=1.5,draw=black,label=left:$c_3$] {};
		\node (a2) at (0:0) [circle,inner sep=1.5,draw=black,label=above:$a_2$] {};
		\path[-] (a1) edge[thick] (b1);
		\path[-] (b1) edge[thick] (ca);
		\path[-] (ca) edge[thick] (b2);
		\path[-] (b2) edge[thick] (c2);
		\path[-] (c2) edge[thick] (a1);
		\path[-] (a1) edge[thick] (b3);
		\path[-] (c2) edge[thick] (b3);
		\path[-] (a1) edge[thick] (c3);
		\path[-] (b1) edge[thick] (c3);
		\path[-] (b1) edge[thick] (a2);
		\path[-] (ca) edge[thick] (a2);
		\path[-] (b2) edge[thick] (a2);
		\path[-] (c2) edge[thick] (a2);
		\end{tikzpicture}
	}
\end{minipage}
\begin{minipage}{1.9in}
	\subfloat[]{\label{figObstructions_3}
		\begin{tikzpicture}[scale=.58]
		\node (c1) at (90:2.5) [circle,inner sep=1.5,draw=black,label=above:$c_1$] {};	
		\node (b1) at (0:2.5) [circle,inner sep=1.5,draw=black,label=right:$b_1$] {};
		\node (c2) at (270:2.5)  [circle,inner sep=1.5,draw=black,label=below:$c_2$] {};
		\node (a1) at (180:2.5) [circle,inner sep=1.5,draw=black,label=left:$a_1$] {};
		\node (a2) at (45:3) [circle,inner sep=1.5, draw=black,label=right:$a_2$] {};
		\node (a3) at (315:3) [circle,inner sep=1.5,draw=black,label=right:$a_3$] {};
		\node (b3) at (225:3) [circle,inner sep=1.5,draw=black,label=left:$b_3$] {};
		\node (b2) at (135:3) [circle,inner sep=1.5,draw=black,label=left:$b_2$] {};
		\node (c3) at (0:0) [circle,inner sep=1.5,draw=black,label=right:$c_3$] {};
		\path[-] (c1) edge[thick] (b1);
		\path[-] (b1) edge[thick] (c2);
		\path[-] (c2) edge[thick] (a1);
		\path[-] (a1) edge[thick] (c1);
		\path[-] (a2) edge[thick,bend right] (b3);
		\path[-] (b3) edge[thick] (c3);
		\path[-] (a2) edge[thick] (c3);
		\path[-] (c1) edge[thick] (a2);
		\path[-] (b1) edge[thick] (a2);
		\path[-] (c1) edge[thick] (b2);
		\path[-] (a1) edge[thick] (b2);
		\path[-] (a1) edge[thick] (b3);
		\path[-] (c2) edge[thick] (b3);
		\path[-] (c2) edge[thick] (a3);
		\path[-] (b1) edge[thick] (a3);
		\end{tikzpicture}
	}
\end{minipage}
\caption{The forbidden sets of edges of the partition intersection graph of three characters that have a perfect phylogeny (adapted from Figure 42 in \cite{Lam2009a}). We note that in \cite{Lam2009a}, there are four forbidden sets of edges, however, one of the sets of edges is a superset of one of the other sets of edges. Thus, only three are needed here.}
\label{figObstructions}
\end{figure}
\end{thm}

\subsection{Solving Three-State Perfect Phylogeny with Two-State Characters}

Here we review a result of Dress and Steel \cite{Dress1992a}. Our exposition closely follows that of \cite{Gusfield2009a}. 

Our goal is to derive a matrix of two-state characters $\overline{M}$ from the matrix $M$ of three-state characters. The properties of $\overline{M}$ are such that they enable use to find a perfect phylogeny for $M$. The matrix $\overline{M}$ contains three characters $c(1)$, $c(2)$, $c(3)$ for each character $c$ in $M$, such that all of the taxa that have state $i$ for $c$ in $M$ are given state 1 for character $c(i)$ in $\overline{M}$, and the other taxa are given state 2 for $c(i)$ in $\overline{M}$.

Since every character in $\overline{M}$ has two states, two characters $c(i)$ and $d(j)$ of $\overline{M}$ are incompatible if and only if the two columns corresponding to $c(i)$ and $d(j)$ contain all four of the pairs $(1,1)$, $(1,2)$, $(2,1)$, and $(2,2)$, otherwise they are compatible. This is known as the {\em four gametes test} \cite{Semple2003a}.

\begin{thm}[Dress and Steel \cite{Dress1992a}]\label{thm3PPto2PP}
There is a perfect phylogeny for $M$ if and only if there is a subset $C$ of the characters of $\overline{M}$ such that 
\begin{itemize}
\item[(i)] the characters in $C$ are pairwise compatible, and
\item[(ii)] for each character $c$ in $M$, $C$ contains at least two of the characters $c(1)$, $c(2)$, $c(3)$.
\end{itemize}
\end{thm} 

Theorem \ref{thm3PPto2PP} was used in \cite{Dress1992a} to give an $O(m^2n)$ time algorithm to decide if there is a perfect phylogeny for $M$. It was also used in \cite{Gusfield2009a} to reduce the three-state perfect phylogeny problem in polynomial time to the well known 2-SAT problem, which is in $P$.

\section{A Simple Characterization of Minimal Obstruction Sets}

In this section, we focus on the case where M is pairwise compatible.  
Our main result is a characterization of the situation where M does not have a perfect phylogeny that is based on the partition intersection graphs for the pairs of characters in M.
Theorem \ref{thmPairCompatible} gives a simple characterization of the situation when $M$ is not pairwise compatible. 

We say that a state $i$ for a character $c$ of $M$ is {\em dependent} precisely when there exists a character $d$ of $M$, and two states $j,k$ of $d$, such that $c(i)$ is incompatible with both $d(j)$ and $d(k)$. The character $d$ is a {\em witness} that state $i$ of $c$ is dependent.

\begin{lem}\label{lemDependentState}
Let $c$ be a character of $M$ and let $i$ be a dependent state of $C$.
Then no pairwise compatible subset of characters in $\overline{M}$ satisfying Theorem \ref{thm3PPto2PP} contains $c(i)$.
\end{lem}
\begin{proof}
Let $I$ be a pairwise compatible subset of the characters in $\overline{M}$ that contains $c(i)$. Since state $i$ of $c$ is dependent, there is a character $b$ in $M$ and two states $j,k$ of $b$, such that $c(i)$ is incompatible with both $b(j)$ and $b(k)$. It follows that $b(j) \not\in I$ and $b(k) \not\in I$. But then $I$ cannot possibly contain two of $b(1)$, $b(2)$, and $b(3)$. Thus, $I$ cannot satisfy the condition required in Theorem \ref{thm3PPto2PP}.
\end{proof}

The next lemma gives a characterization of when a state is dependent using partition intersection graphs. We first introduce some notation: if $p : p_1p_2p_3p_4p_5$ is a path of length four in a graph, then we write $\mathrm{middle}[p]$ to denote $p_3$, the {\em middle} vertex of $p$.

\begin{lem}\label{lemPathDependent}
Let $M$ be pairwise compatible. A state $i$ of a character $c$ of $M$ is a dependent state if and only if there is a character $d$ of $M$ and a path $p$ of length four in $\pig(M[c,d])$ with $\mathrm{middle}[p]=c_i$. 
\end{lem}

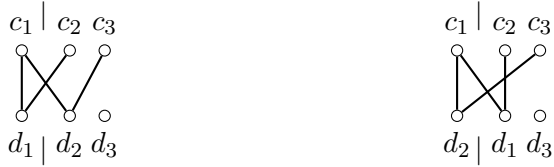
\begin{figure}[!b]
\vspace{-10pt}
\centering
\subfloat[$c(1)$ and $d(1)$ are incompatible.]{
	\hspace{45pt}
	\begin{tikzpicture}[scale=.58]
		\node (c1) at (0,0) [circle,inner sep=1.5,draw=black,label=above:$c_1$] {};
		\node (c1bar) at (.5,0) [draw=none,label=above:$|$] {};
		\node (c2) at (1.1,0) [circle,inner sep=1.5,draw=black,label=above:$c_2$] {};
		\node (c3) at (1.9,0) [circle,inner sep=1.5,draw=black,label=above:$c_3$] {};
		\node (d1) at (0,-1.5) [circle,inner sep=1.5,draw=black,label=below:$d_1$] {};
		\node (d1bar) at (.5,-1.5) [draw=none,label=below:$|$] {};
		\node (d2) at (1.1,-1.5) [circle,inner sep=1.5,draw=black,label=below:$d_2$] {};
		\node (d3) at (1.9,-1.5) [circle,inner sep=1.5,draw=black,label=below:$d_3$] {};
		\path[-] (c2) edge[thick] (d1);
		\path[-] (d1) edge[thick] (c1);
		\path[-] (c1) edge[thick] (d2);
		\path[-] (d2) edge[thick] (c3);
	\end{tikzpicture}
	\hspace{45pt}
}
\quad
\subfloat[$c(1)$ and $d(2)$ are incompatible.]{
	\hspace{45pt}
	\begin{tikzpicture}[scale=.58]
		\node (c1) at (0,0) [circle,inner sep=1.5,draw=black,label=above:$c_1$] {};
		\node (c1bar) at (.5,0) [draw=none,label=above:$|$] {};
		\node (c2) at (1.1,0) [circle,inner sep=1.5,draw=black,label=above:$c_2$] {};
		\node (c3) at (1.9,0) [circle,inner sep=1.5,draw=black,label=above:$c_3$] {};
		\node (d2) at (0,-1.5) [circle,inner sep=1.5,draw=black,label=below:$d_2$] {};
		\node (d2bar) at (.5,-1.5) [draw=none,label=below:$|$] {};
		\node (d1) at (1.1,-1.5) [circle,inner sep=1.5,draw=black,label=below:$d_1$] {};
		\node (d3) at (1.9,-1.5) [circle,inner sep=1.5,draw=black,label=below:$d_3$] {};
		
		\path[-] (c2) edge[thick] (d1);
		\path[-] (d1) edge[thick] (c1);
		\path[-] (c1) edge[thick] (d2);
		\path[-] (d2) edge[thick] (c3);
	\end{tikzpicture}
	\hspace{45pt}
}
\caption{Illustrating the proof of Lemma \ref{lemPathDependent}.}
\label{figPathIncompatible}
\end{figure}

\begin{proof}
W.l.o.g. assume that $i=1$, i.e., $c_i=c_1$.

($\Rightarrow$)
Since $1$ is a dependent state of $c$, there exists a character $d$ in $M$ such that $c(1)$ is incompatible with two of $d(1)$, $d(2)$, and $d(3)$. W.l.o.g., assume $c(1)$ is incompatible with both $d(1)$ and $d(2)$. Then, $c_1d_1$ and $c_1d_2$ are edges of $\pig(M[c,d])$, and, since $M$ has no cycles, either $d_2c_2$ and $d_1c_3$ or $d_2c_3$ and $d_1c_2$ are edges of $G$. If $d_2c_2$ and $d_1c_3$ are edges of $\pig(M[c,d])$, then $c_2d_2c_1d_1c_3$ is the required path of length four. If $d_2c_3$ and $d_1c_2$ are edges of $\pig(M[c,d])$, then $c_3d_2c_1d_1c_2$ is the required path of length four.

($\Leftarrow$)
Let $d$ be a character of $M$ such that there is a path $p$ of length four in $\pig(M[c,d])$ with $\mathrm{middle}[p]=c_1$.
Since $\pig(M[c,d])$ cannot contain edges between to states of the same character, we can assume w.l.o.g. that $p$ is the path $c_2d_1c_1d_2c_3$.
Then, it is easy to verify that $c(1)$ is incompatible with both $d(1)$ and $d(2)$. This is illustrated in Figure \ref{figPathIncompatible}.
\end{proof}

\begin{lem}\label{lemDependentPP}
If $M$ is pairwise compatible and there is a character $c$ of $M$ that has two dependent states, then no perfect phylogeny exists for $M$.
\end{lem}
\begin{proof}
Let $i$ and $j$ be two dependent states of $c$. 
Then, by Lemma \ref{lemDependentState}, no pairwise compatible subset $I$ of the characters of $\overline{M}$ that satisfy the condition required in Theorem \ref{thm3PPto2PP} can contain $c(i)$ or $c(j)$.
But then $I$ can only contain one of $c(1)$, $c(2)$, or $c(3)$. Hence, no pairwise compatible subset $I$ of the characters of $\overline{M}$ can satisfy the condition required in Theorem \ref{thm3PPto2PP}. Hence, by Theorem  \ref{thm3PPto2PP}, there is no perfect phylogeny for $M$.
\end{proof}

We now show that the converse of Lemma \ref{lemDependentPP} holds.

\begin{lem}\label{lemPPDependent}
If $M$ is pairwise compatible and has no perfect phylogeny, then there exists a character $c$ of $M$ that has two dependent states.
\end{lem}
\begin{proof}
By Theorem \ref{thm3StateGametes}, there exists characters $a,b,c$ in $M$ such that $G=\pig(M[a,b,c])$ (under possibly renaming of states) contains all of the edges of at least one of the subgraphs of Figure \ref{figObstructions}. If $G$ contains all of the edges of Figure \ref{figObstructions_1}, then $c_3b_1c_1b_2c_2$ is a path witnessing that $c_1$ is dependent and $c_3a_1c_2a_3c_1$ is a path witnessing that $c_2$ is dependent (this is illustrated in Figure \ref{figPig1Bolded}).
If $G$ contains all of the edges of Figure \ref{figObstructions_2}, then $c_3b_1c_1b_2c_2$ is a path witnessing that $c_1$ is dependent and $c_3a_1c_2a_2c_1$ is a path witnessing that $c_2$ is dependent (this is illustrated in Figure \ref{figPig2Bolded}).
If $G$ contains all of the edges of Figure \ref{figObstructions_2}, then $c_3a_2c_1a_1c_2$ is a path witnessing that $c_1$ is dependent  and $c_3b_3c_2b_1c_1$ is a path witnessing that $c_2$ is dependent (this is illustrated in Figure \ref{figPig3Bolded}).
In all three cases, $M$ contains a character that has two dependent states.
\end{proof}
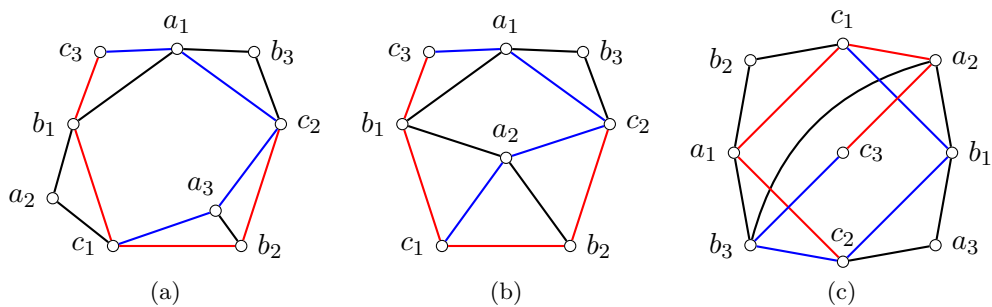
\begin{figure}[!h]
	\centering
	\vspace{-10pt}
	\subfloat[]{\label{figPig1Bolded}
		\begin{tikzpicture}[scale=.58]
		\node (a1) at (90:2.5) [circle,inner sep=1.5,draw=black,label=above:$a_1$] {};
		\node (b1) at (162:2.5) [circle,inner sep=1.5,draw=black,label=left:$b_1$] {};
		\node (c1) at (234:2.5) [circle,inner sep=1.5,draw=black,label=left:$c_1$] {};
		\node (b2) at (306:2.5)  [circle,inner sep=1.5,draw=black,label=right:$b_2$] {};	
		\node (c2) at (18:2.5)  [circle,inner sep=1.5,draw=black,label=right:$c_2$] {};
		\node (b3) at (54:3) [circle,inner sep=1.5,draw=black,label=right:$b_3$] {};
		\node (c3) at (126:3) [circle,inner sep=1.5,draw=black,label=left:$c_3$] {};
		\node (a2) at (198:3) [circle,inner sep=1.5,draw=black,label=left:$a_2$] {};
		\node (a3) at (306:1.5) [circle,inner sep=1.5,draw=black,label=above:$a_3$\ \ \ \ ] {};
		\path[-] (a1) edge[thick] (b1);
		\path[-] (b1) edge[thick,red] (c1);
		\path[-] (c1) edge[thick,red] (b2);
		\path[-] (b2) edge[thick,red] (c2);
		\path[-] (c2) edge[blue,thick] (a1);
		\path[-] (a1) edge[thick] (b3);
		\path[-] (c2) edge[thick] (b3);
		\path[-] (a1) edge[thick,blue] (c3);
		\path[-] (b1) edge[thick,red] (c3);
		\path[-] (b1) edge[thick] (a2);
		\path[-] (c1) edge[thick] (a2);
		\path[-] (c1) edge[blue,thick] (a3);
		\path[-] (b2) edge[thick] (a3);
		\path[-] (c2) edge[blue,thick] (a3);
		\end{tikzpicture}
	}
	\subfloat[]{\label{figPig2Bolded}
		\begin{tikzpicture}[scale=.58]
		\node (a1) at (90:2.5) [circle,inner sep=1.5,draw=black,label=above:$a_1$] {};
		\node (b1) at (162:2.5) [circle,inner sep=1.5,draw=black,label=left:$b_1$] {};
		\node (c1) at (234:2.5) [circle,inner sep=1.5,draw=black,label=left:$c_1$] {};
		\node (b2) at (306:2.5)  [circle,inner sep=1.5,draw=black,label=right:$b_2$] {};	
		\node (c2) at (18:2.5)  [circle,inner sep=1.5,draw=black,label=right:$c_2$] {};
		\node (b3) at (54:3) [circle,inner sep=1.5,draw=black,label=right:$b_3$] {};
		\node (c3) at (126:3) [circle,inner sep=1.5,draw=black,label=left:$c_3$] {};
		\node (a2) at (0:0) [circle,inner sep=1.5,draw=black,label=above:$a_2$] {};
		\path[-] (a1) edge[thick] (b1);
		\path[-] (b1) edge[thick,red] (c1);
		\path[-] (c1) edge[thick,red] (b2);
		\path[-] (b2) edge[thick,red] (c2);
		\path[-] (c2) edge[thick,blue] (a1);
		\path[-] (a1) edge[thick] (b3);
		\path[-] (c2) edge[thick] (b3);
		\path[-] (a1) edge[thick,blue] (c3);
		\path[-] (b1) edge[thick,red] (c3);
		\path[-] (b1) edge[thick] (a2);
		\path[-] (c1) edge[thick,blue] (a2);
		\path[-] (b2) edge[thick] (a2);
		\path[-] (c2) edge[thick,blue] (a2);
		\end{tikzpicture}
	}
	\subfloat[]{\label{figPig3Bolded}
		\begin{tikzpicture}[scale=.58]
		\node (c1) at (90:2.5) [circle,inner sep=1.5,draw=black,label=above:$c_1$] {};	
		\node (b1) at (0:2.5) [circle,inner sep=1.5,draw=black,label=right:$b_1$] {};
		\node (c2) at (270:2.5)  [circle,inner sep=1.5,draw=black,label=above:$c_2$] {};
		\node (a1) at (180:2.5) [circle,inner sep=1.5,draw=black,label=left:$a_1$] {};
		\node (a2) at (45:3) [circle,inner sep=1.5, draw=black,label=right:$a_2$] {};
		\node (a3) at (315:3) [circle,inner sep=1.5,draw=black,label=right:$a_3$] {};
		\node (b3) at (225:3) [circle,inner sep=1.5,draw=black,label=left:$b_3$] {};
		\node (b2) at (135:3) [circle,inner sep=1.5,draw=black,label=left:$b_2$] {};
		\node (c3) at (0:0) [circle,inner sep=1.5,draw=black,label=right:$c_3$] {};
		\path[-] (a2) edge[thick,bend right] (b3);
		\path[-] (c1) edge[thick,blue] (b1);
		\path[-] (b1) edge[thick,blue] (c2);
		\path[-] (c2) edge[thick,red] (a1);
		\path[-] (a1) edge[thick,red] (c1);
		\path[-] (b3) edge[thick,blue] (c3);
		\path[-] (a2) edge[thick,red] (c3);
		\path[-] (c1) edge[thick,red] (a2);
		\path[-] (b1) edge[thick] (a2);
		\path[-] (c1) edge[thick] (b2);
		\path[-] (a1) edge[thick] (b2);
		\path[-] (a1) edge[thick] (b3);
		\path[-] (c2) edge[thick,blue] (b3);
		\path[-] (c2) edge[thick] (a3);
		\path[-] (b1) edge[thick] (a3);
		\end{tikzpicture}
	}
	\caption{Illustrating the proof of Lemma \ref{lemPPDependent}.}
\end{figure}

Lemmas \ref{lemDependentPP} and \ref{lemPPDependent} together immediately imply our main theorem.

\begin{thm}\label{thmMain}
If $M$ is pairwise compatible, then there is a perfect phylogeny for $M$ if and only if there is at most one dependent state of each character $c$ of $M$.
\end{thm}

\begin{obs}\label{obsObstructionSet}
Let $M$ be pairwise compatible and let $c$ be a character of $M$ with two dependent states. Let $a$ be a witness for one dependent state of $c$ and let $b$ be a witness for another dependent state of $c$. Then, the set $\{a,b,c\}$ is an obstruction set for $M$. 
\end{obs}

This leads to the following $O(m^2n)$ time algorithm to find a minimal obstruction set for $M$, if one exists.\\

\begin{algorithm}[!h]
\caption{MinimalObstructionSet($M$)}
\begin{algorithmic}[1]
\REQUIRE $M$ is an $n$ by $m$ matrix of integers from the set $\{1,2,3\}$.
\ENSURE A minimal obstruction set for $M$ if one exists, otherwise the empty set.
\FORALL{characters $x$ in $M$}
	\FORALL{states $i$ of $x$}
		\STATE $\mathrm{mark}[x_i] \leftarrow \emptyset$;
	\ENDFOR
\ENDFOR
\STATE $S \leftarrow \emptyset$;
\FORALL{pairs of characters $a,b$ in $M$}
	\STATE $G \leftarrow \pig(M[a,b])$;
	\IF{$G$ contains a cycle}
		\RETURN $\{a,b\}$;
	\ELSIF{$S = \emptyset$}
		\FORALL{$x_i  \in \{a_1,a_2,a_3,b_1,b_2,b_3\}$ such that $\mathrm{mark}[x_i]$ is empty}
			\IF{there is a path $p$ of length four in $G$ with $\mathrm{middle}[p]=c_i$}
				\STATE $\mathrm{mark}[x_i] \leftarrow \{a,b\} \setminus \{x\}$;
			\ENDIF
		\ENDFOR
		\IF{two states $i,j$ of $a$ have non-empty marks}
			\STATE $S \leftarrow \{a\} \cup \mathrm{mark}[a_i] \cup \mathrm{mark}[a_j]$;
		\ELSIF{two states $i,j$ of $b$ have non-empty marks}
			\STATE $S \leftarrow \{b\} \cup \mathrm{mark}[b_i] \cup \mathrm{mark}[b_j]$;
		\ENDIF
	\ENDIF
\ENDFOR
\RETURN $S$;
\end{algorithmic}
\end{algorithm}
\vspace{6pt}

The correctness of the algorithm follows from Theorem \ref{thmPairCompatible}, Theorem \ref{thmMain}, Observation \ref{obsObstructionSet}, and Lemma \ref{lemPathDependent}.
To see that the algorithm takes $O(m^2n)$ time note that the runtime is dominated by the loop of lines 5-16 which executes once for each of the $O(m^2)$ pairs of characters in $M$. Constructing the partition intersection graph of two three-state characters takes $O(n)$ time. Since the partition intersection graph of two three-state characters is of constant size, each of the other operations performed in the loop take constant time.

We note that if no obstruction set exists for $M$, then a perfect phylogeny for $M$ can be constructed in $O(m^2n)$ time by using one of the existing algorithms for the three-state perfect phylogeny problem  \cite{Dress1992a, Kannan1994a, Agarwala1994a, Kannan1997a, Gusfield2009a}.

\section*{Acknowledgments}
This work was supported in part by the National Science Foundation under grants CCF-1017189 and DEB-0829674.

{\small
\bibliographystyle{abbrv}
\bibliography{../Bibs/PhyloBib/PhyloBib}
}
\end{document}